\documentclass[12pt,preprint]{aastex}

\received{}
\accepted{}
\journalid{}{}
\articleid{}{}
\paperid{}
\cpright{AAS}{2002}
\ccc{}

\slugcomment{To be submitted to the {\it Astrophysical Journal}}

\shorttitle{The Molecular Continuum Opacity of $^{24}$MgH} 
\shortauthors{P. F. Weck et al.}

\begin{document}

\title{The Molecular Continuum Opacity of $^{24}$MgH in Cool
            Stellar Atmospheres}

\author{P. F. Weck, A. Schweitzer, P. C. Stancil, \& P. H. Hauschildt} 
\affil{Department of Physics and Astronomy and Center for Simulational 
Physics, \\
The University of Georgia, Athens, GA 30602-2451}
\email{weck-,andy-,stancil-,yeti@physast.uga.edu}

\author{K. Kirby}
\affil{ITAMP, Harvard-Smithsonian Center for Astrophysics\\
       60 Garden St., Cambridge, MA 02138}
\email{kirby@cfa.harvard.edu}

\begin{abstract}
The opacity due to photodissociation of $^{24}$MgH is investigated
in the atmospheres of cool stars. The lowest two electronic transitions
$A~^2\Pi\leftarrow X~^2\Sigma^+$ and $B'~^2\Sigma^+\leftarrow X~^2\Sigma^+$
are considered where the cross sections for the latter were published
previously (Weck, Stancil, \& Kirby 2002a) while the former are 
presented in this work.  
Model atmospheres calculated with the {\tt PHOENIX} code are
used to investigate the effect of the photodissociation opacity on
spectra of cool stars.
The $A~^2\Pi\leftarrow X~^2\Sigma^+$
photodissociation cross sections are obtained  
using a combination of {\it ab initio} and experimentally derived
potential curves and dipole transition moments. 
Partial cross sections have been evaluated 
over the accessible wavelength range $\lambda\lambda~1770-4560$~\AA~for 
all rotational transitions from the vibrational levels $v''=0-11$.
Assuming a Boltzmann distribution of the
rovibrational levels of the $X~^2\Sigma^+$ state, LTE photodissociation
cross sections are presented for temperatures between 1000 and
5000~K. Shape resonances, arising from rotational 
predissociation of quasi-bound levels of the $A~^2\Pi$ state near 
threshold, characterize the LTE photodissociation cross sections. 
A sum rule is proposed as a check on the accuracy of the 
photodissociation calculations.    
\end{abstract}


\section{Introduction}

The lack of accurate and complete molecular line and continuum opacity 
data has been a serious limitation to developing atmospheric models of 
cool stars (M and later), solar system planets, and Extrasolar 
Giant Planets (EGPs). 
Sophisticated modeling programs, such as {\tt PHOENIX} \citep{hau99}, 
require high quality opacity data in order to produce synthetic spectra 
and predict physical parameters (e.g., surface chemical composition, effective 
temperature, etc). Typically, atmosphere models include molecular bands 
with hundreds of millions of spectral lines, mostly derived from molecular 
band or Hamiltonian models. Moreover, atmospheric models do not
consider the effect of molecular 
photodissociation processes, which may play a role in the opacity at 
visible and UV wavelengths. An exception is the work of \citet{kur87}
who included LTE photodissociation calculations of CH and OH in solar
and cool stellar atmosphere models. Further, it has been suggested
by \citet{sho94,sho96} that a missing opacity source between 2500 and
4000 \AA~ might be due to MgH, based on the ground rovibrational
photodissociation cross section of \cite{kir79} which peaks at
2920 \AA~, but has a threshold at 3100 \AA. However, the LTE
$B'~^2\Sigma^+ \leftarrow X~^2\Sigma^+$
MgH calculations of \citet*{wec02a} find that the cross sections have
significant amplitude to nearly 4500 \AA.     

In the present work, we have performed extensive calculations of 
rovibrationally-resolved photodissociation cross sections of $^{24}$MgH through 
the $A~^2\Pi \leftarrow X~^2\Sigma^+$ transition, using the most accurate 
available molecular data. Calculations were performed for 
the full range of 313 rovibrational levels $(v'',J'')$ in the ground electronic 
state. Assuming a Boltzmann distribution of the
rovibrational levels of the $X~^2\Sigma^+$ state, LTE photodissociation
cross sections are also presented for temperatures between 1000 and
5000~K. These photodissociation cross sections as well as those
of \citet{wec02a} for the $B'~^2\Sigma^+ \leftarrow X~^2\Sigma^+$
transition have been included in cool stellar atmosphere calculations
to test if MgH photodissociation is the missing opacity postulated by 
\citet{sho94,sho96}.
In this paper we investigate the importance of the bound--free
opacity in both very cool dwarfs and the objects discussed in
\citet{sho94,sho96} (K type giants and solar type dwarfs).

Atomic units are used throughout to discuss the molecular
calculations unless otherwise stated.

\section{Theory of Photodissociation} \label{theo}

Potential curves of the $X~^2\Sigma^+$ 
and $A~^2\Pi$ electronic states of MgH 
calculated by Saxon, Kirby \& Liu (1978), 
as well as the connecting dipole transition moment, 
were used in the present study. 
Shifts have been applied to the potential curves, in order to correct 
for discrepancies between the theoretical dissociation energy 
and the experimental values given by \cite{bal76} and 
\cite{bal78}. Details about this procedure, together with the fits used for 
short- and long-range internuclear separations, have 
been described in a previous 
publication \citep{wec02b}.      

For absorption into the 
rovibrational continuum of electronic state $f$ from the rovibrational
state $v''J''$ of electronic state $i$, 
the partial rotational photodissociation cross section can be written as 
\citep{wec02a}
\begin{equation} \label{pcs} 
\sigma_{v''J''} = 2.69\times 10^{-18}~\Delta E_{k'J',v''J''}
\frac {1}{2J''+1}\sum_{J'=J''-1}^{J''+1}S_{J'} 
\vert D^{fi}_{k'J',v''J''}\vert^2~{\rm cm}^2, 
\end{equation}
where $\Delta E_{k'J',v''J''}$ is the energy of the absorbed photon 
and $S_{J'}$ are the H\"{o}nl-London factors defined, 
for the $A~^2\Pi \leftarrow X~^2\Sigma^+$ transition as 
\begin{eqnarray}\label{hlfac}
S_{J'}(J'')&=&
\left\{
\begin{array}{ll}
(J''-1)/2,&~J'=J''-1~(\mbox{P branch})\\
(2J''+1)/2,&~J'=J''~~~~~~(\mbox{Q branch})\\
(J''+2)/2,&~J'=J''+1~(\mbox{R branch})
\end{array} \right .   
\end{eqnarray} 
according to the 
convention proposed by \citet{whi74}.
In equation (\ref{pcs}), the term 
$D^{fi}_{k'J',v''J''}=<\chi_{k'J'}\vert D^{fi}\vert\chi_{v''J''}>$ 
is the matrix element of the electric dipole transition moment responsible for 
absorption from the $i$ into the $f$ electronic state. 
The continuum wave functions $\chi_{k'J'}(R)$ are normalized 
such that they behave asymptotically as
\begin{equation}
\chi_{k'J'}(R)\sim\sin(k'R-\case{\pi}{2}J' +\delta_{J'}).  
\end{equation}

The continuum and bound wave functions $\chi_{k'J'}$ and $\chi_{v''J''}$, 
respectively, were obtained by solution of the radial 
Schr\"{o}dinger equation for the $A$ and $X$ states, respectively, using 
a standard numerical method \citep{coo61}. 
Wave functions were obtained on a grid with a stepsize of 
$1\times10^{-3}~\mbox{a}_o$ over the internuclear distance range 
$0.5 \leqslant R \leqslant 200~\mbox{a}_o$.

\section{Photodissociation Calculations}

\begin{figure}
\plottwo{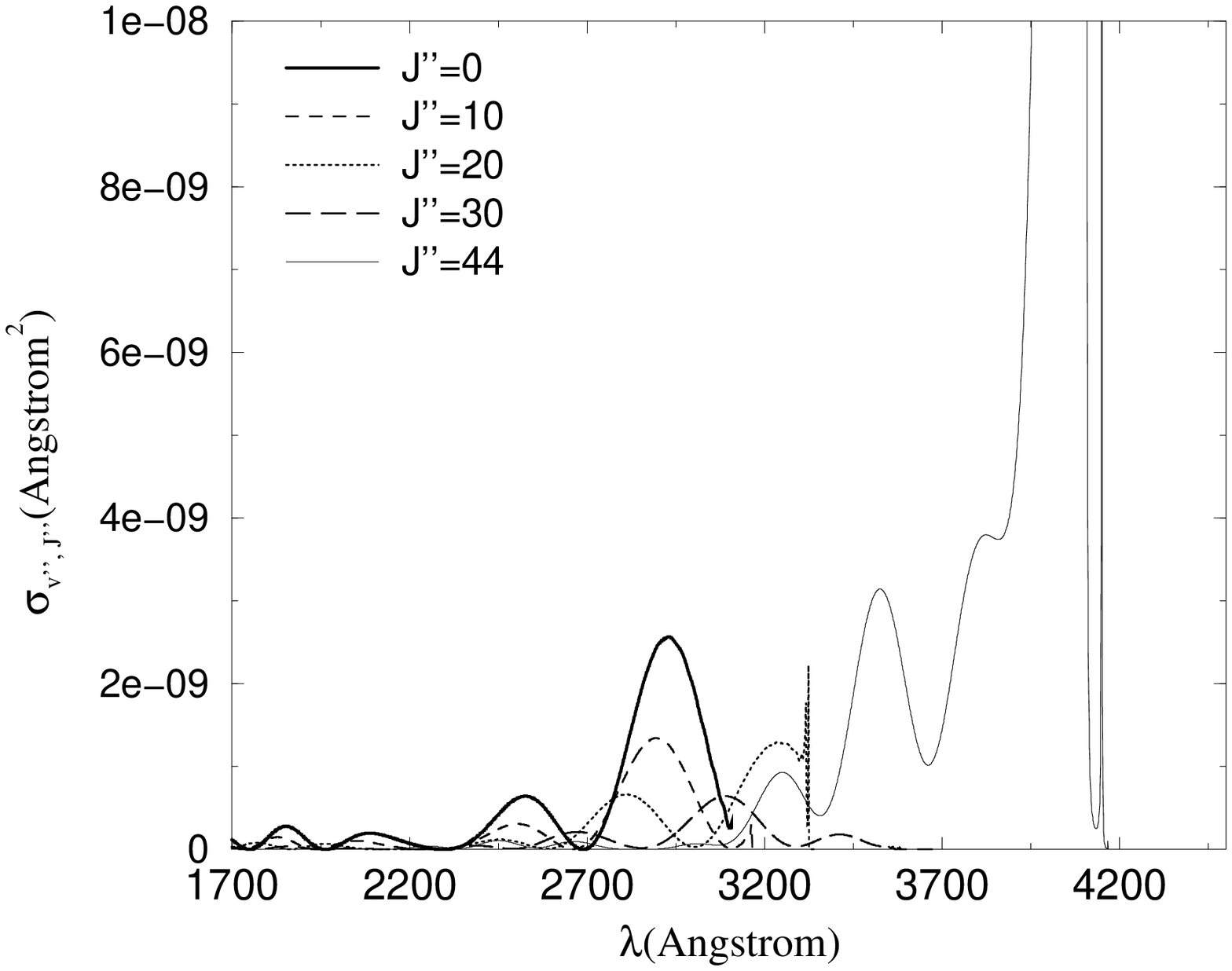}{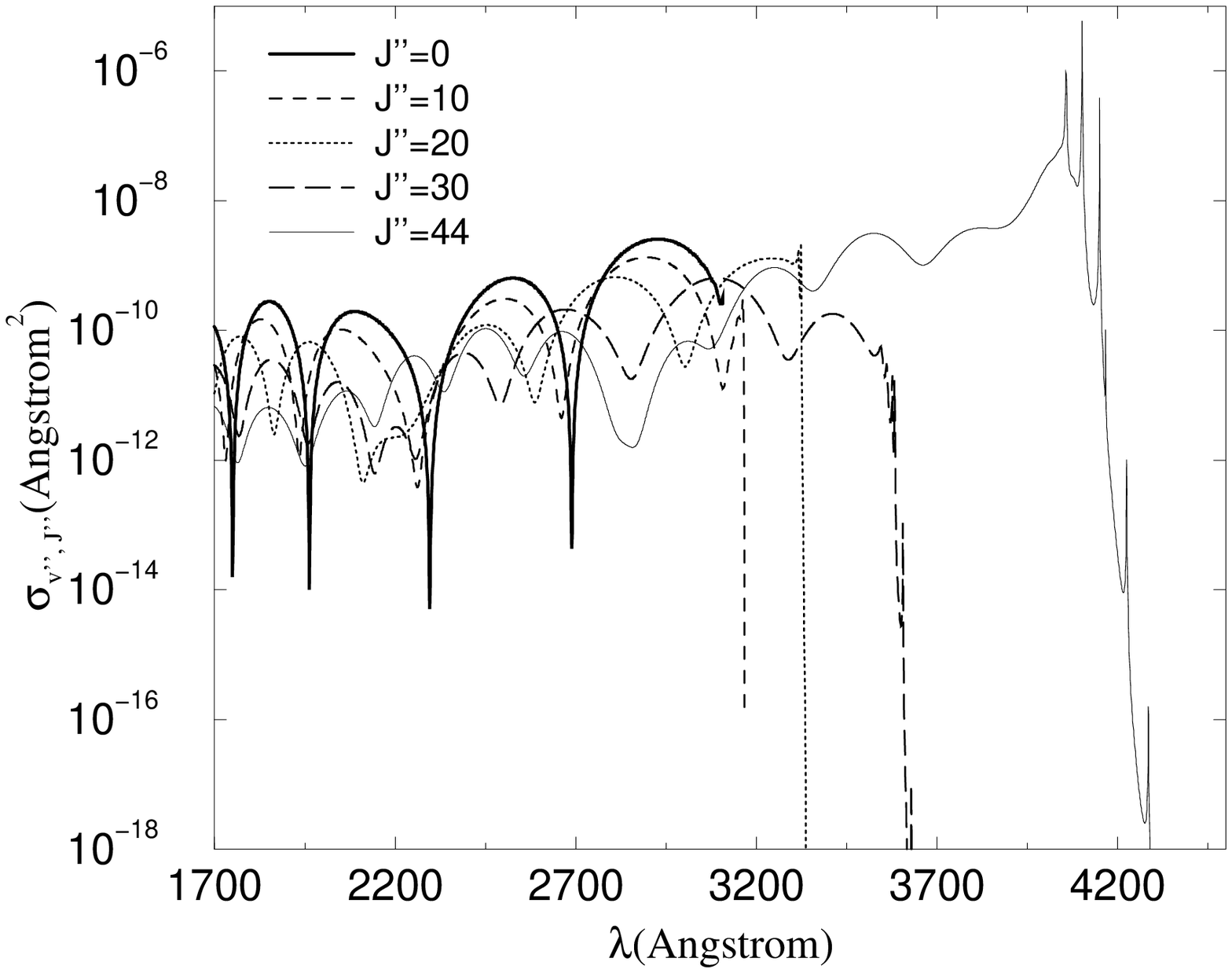}
\caption{Partial cross sections $\sigma_{v''J''}$ for transitions from 
the ground vibrational level 
$v''=0$ with the rotational quantum numbers $J''=0$, 10, 20, 30, and 44 of 
the $X~^2\Sigma^+$ electronic state of $^{24}$MgH. 
On the left: linear scale; on the right:
logarithmic scale.
\label{fig1}}
\end{figure}

\begin{figure}
\plotone{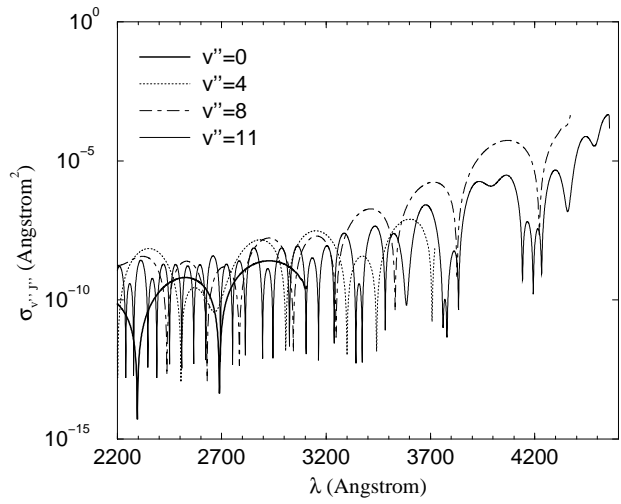}
\caption{Partial cross sections $\sigma_{v''J''}$ for transitions from 
vibrational levels $v''=0$, 4, 8, and 11 with the rotational quantum number 
$J''=0$ of the $X~^2\Sigma^+$ electronic state of $^{24}$MgH. 
\label{fig2}}
\end{figure}

\begin{figure}
\plotone{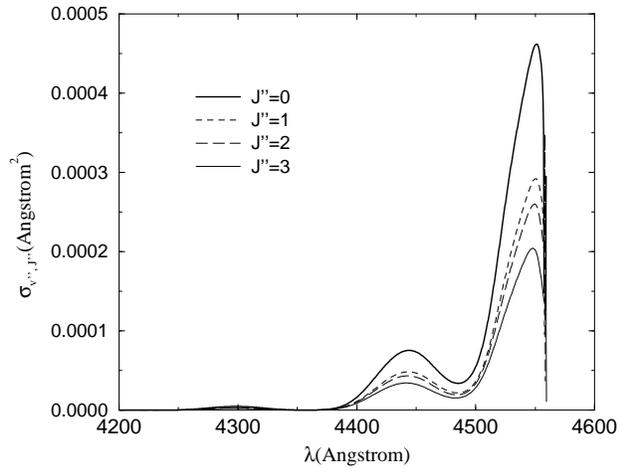}
\caption{Partial cross sections $\sigma_{v''J''}$ for transitions from 
the vibrational level 
$v''=11$ with the rotational quantum numbers $J''=0$, 1, 2, and 3 of 
the $X~^2\Sigma^+$ electronic state of $^{24}$MgH. 
\label{fig3}}
\end{figure}

In Figures \ref{fig1}-\ref{fig3} are shown some representative 
rovibrationally-resolved cross sections for photodissociation through the
$A~^2\Pi\leftarrow X~^2\Sigma^+$ transition\footnote{The complete 
list of data for $A\leftarrow X$ photodissociation 
cross sections is available online 
at the UGA Molecular Opacity Project Database 
website \url{http://www.physast.uga.edu/ugamop/}}.  

Figure \ref{fig1} presents our results for the partial cross sections 
$\sigma_{v''J''}$, as a function of wavelength, for the transitions from 
the vibrational level 
$v''=0$ with the rotational quantum numbers $J''=0$, 10, 20, 30 and 44 of 
the $X~^2\Sigma^+$ electronic state. From the quantitative point of view, 
cross sections are seen to be small, if compared to the values obtained 
for photodissociation through the $B'~^2\Sigma^+\leftarrow X~^2\Sigma^+$ which 
are several orders of magnitude larger for the same wavelength range \citep{wec02a}. 
This could be expected since the bound states of the $A~^2\Pi$ and $X~^2\Sigma^+$ 
electronic states largely overlap, i.e., the sum of the Franck-Condon factors 
over all the upper vibrational levels $v'$ is close to unity as
originally pointed out by \citet{kir79}.
Since the photodissociation cross sections are very small, a 
sum rule with rotational resolution has also been proposed as a check on 
the accuracy of our numerical calculations and is presented in the Appendix.
On the other hand, 
some interesting qualitative features are observed, such as nearly periodic 
nodes and antinodes. This reflects the strongly oscillating behaviour 
of the continuum wave functions, since the $A\leftarrow X$ dipole transition 
moment decreases monotonically to zero as the internuclear separation increases. 
It is also worth noting the presence of shape resonances generated by 
predissociation of quasi-bound rotational levels near threshold. The magnitude 
of such resonances can reach several decades, as can be seen for $J''=44$, 
the topmost rotational level for $v''=0$. 
 
Partial cross sections for transitions from vibrational levels $v''=0$, 4, 8 
and 11 with the rotational quantum number $J''=0$ of the $X~^2\Sigma^+$ 
electronic state are presented in Figure \ref{fig2}. As can be seen in the 
previous figure, the threshold limit extends to larger wavelengths as  
$v''$ and/or $J''$ increase, corresponding to a decrease in the amount 
of energy necessary to reach the dissociation limit. 
In this connection, it is worth mentioning that the cross section value at threshold 
increases significantly with the vibrational quantum number $v''$. This may be 
explained by the fact that the broad maximum of the continuum 
wave function near the left 
turning point overlaps with the dipole transition moment over a wider 
range of internuclear distances.

Figure \ref{fig3} shows the partial cross sections for transitions from 
the topmost vibrational level $v''=11$ with the rotational quantum numbers 
$J''=0$, 1, 2 and 3 of the $X~^2\Sigma^+$ electronic state. Few qualitative 
differences are observed between the maximum rotational number $J''=3$ and 
$J''=0$. In fact, in the case of such high-lying vibrational states, the spacing 
between all the rotational levels is small, thus resulting in few diferences between 
their respective wave functions which sample almost the same potential region.  

\begin{figure}
\plottwo{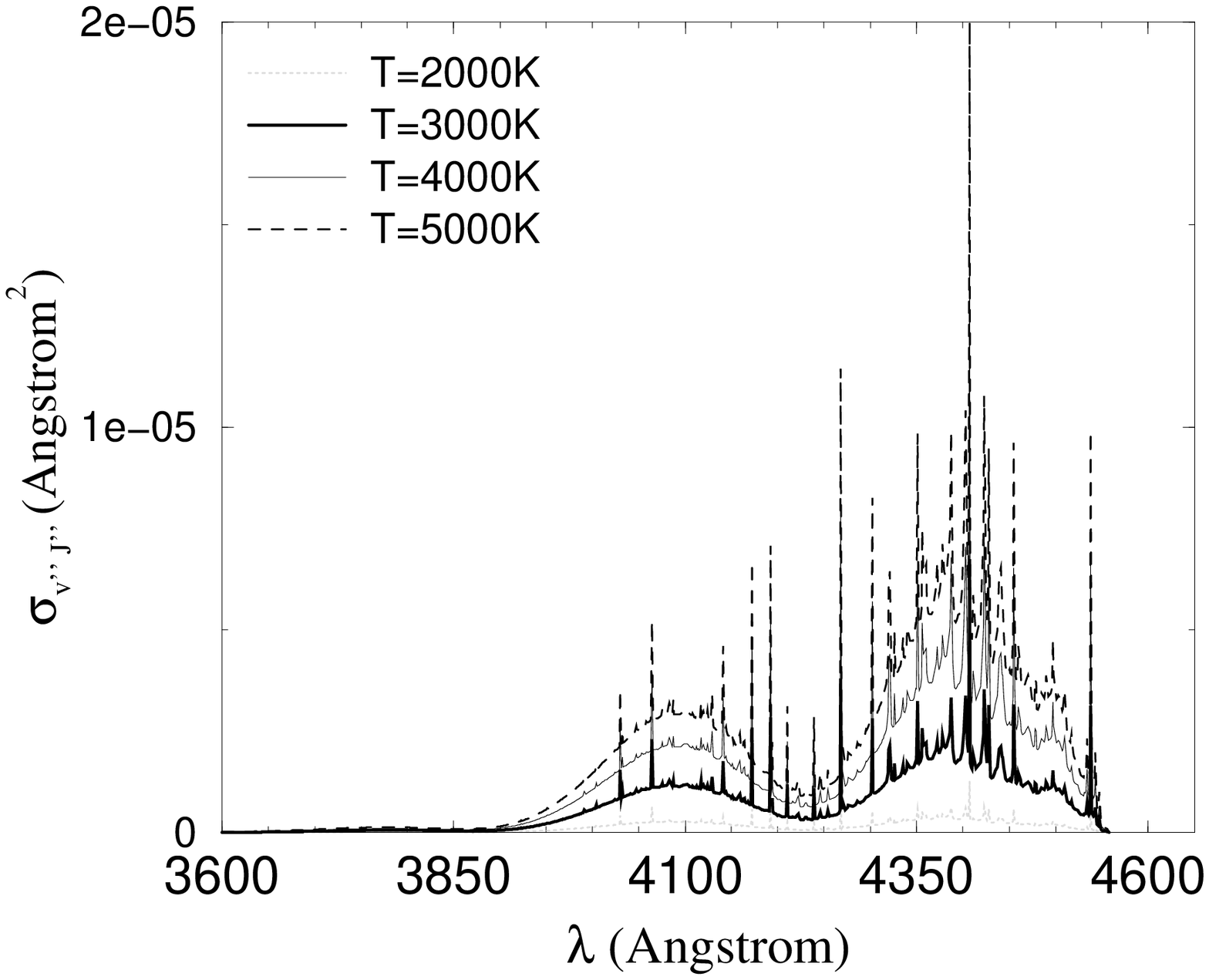}{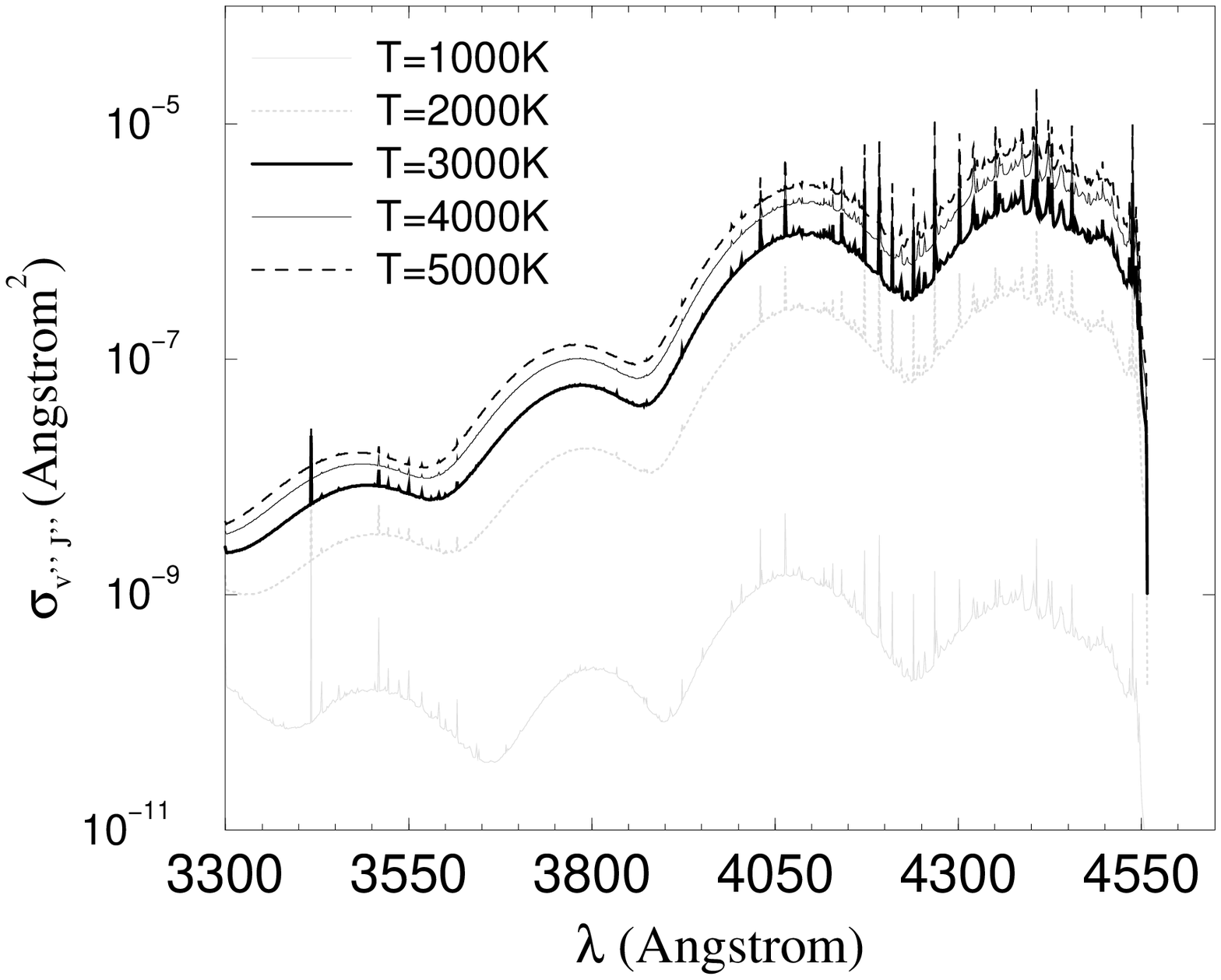}
\caption{The $^{24}$MgH LTE photodissociation cross section
as a function of wavelength for temperatures between $T=1000$ and 5000~K. 
On the left: linear scale; on the right: logarithmic scale.
\label{fig4}}
\end{figure}

Assuming a Boltzmann population distribution of the rovibrational levels of 
the electronic ground state $X~^2\Sigma^+$, the LTE photodissociation cross 
section as a function of the wavelength is shown in Figure \ref{fig4} 
for temperatures between $T=1000$ and 5000 K. 
The LTE cross section shows strong shape resonances, arising from rotational 
predissociation of quasi-bound levels near thresholds, as was observed previously 
for the partial cross sections represented in Figure \ref{fig1}.
The cross section exhibits two broad peaks centered around $\lambda=4100$~and 
$4400$ \AA,~respectively. Two broad peaks were also observed for
the $B'\leftarrow X$ LTE cross sections, but centered near 2780 \AA~
and $\sim$3800 \AA~, with a mininum at 3200 \AA~ \citep{wec02a}. 
For the $A\leftarrow X$
transition, the relative minimum observed around 
$\lambda=4200$ \AA~appears to be the result of a node common to most of 
the high-lying vibrational states, whose partial cross sections strongly 
contribute to the LTE photodissociation cross section. This feature is clearly 
illustrated in Figure \ref{fig2} for the vibrational levels $v''=8$ and $11$ 
on the wavelength range $\lambda\lambda=3800-4600$ \AA, where the LTE cross 
section shows maxima.
Let us note that the LTE cross sections increase with temperature, as 
expected, since more high-lying rovibrational states with lower thresholds are 
occupied at high temperature.    


\subsection{Atmosphere Models}

The  models  used for this  work were calculated as described in
\cite{LimDust}.  These models  and their comparisons to earlier versions  are
the subject of a  separate publication \cite[]{LimDust} and we thus do not
repeat the detailed description of the models here.  However, we will briefly
summarize the major physical properties.  The models are based on the Ames
H$_2$O  and TiO line lists by \cite{ames-water-new} and \cite{ames-tio} and
also include as a new addition the linelists for FeH by \cite{FeHberk2} and for
VO and CrH by  R.  Freedman  (NASA-Ames,  private communication).  The models
account for equilibrium formation of dust and condensates and include grain
opacities for 40 species.  In the following, the models will be referred to as
``AMES-dusty'' for models in which the dust particles stay in the layers in
which they have formed and ``AMES-cond'' for models in which the dust particles
have sunk below the atmosphere from the layers in which they originally formed.
We  stress  that large uncertainties persist  in the water opacities for parts
of the  temperature range of this work \citep{2000ApJ...539..366A}. However,
the MgH bound-free transitions are in the optical 
and only slightly affected by the
quality of the water opacities.

In addition to the opacities sources listed above and in \citet[and references
therein]{LimDust} we added the new photodissociation opacity sources from this
work and from \cite{wec02a} to our opacity
database.  In order to assess the effects of the new MgH photodissociation
data, we compare spectra calculated with and without these opacity sources.  The
original AMES grid was calculated for effective temperatures of M, L and
T~dwrafs. The hotter models in this work are based on the same physics as the
AMES grid and merely differ in the effective temperature.  The models used in
the following discussion were all iterated to convergence for the parameters
indicated.  The high resolution spectra which have the individual opacity
sources  selected are calculated on top of the models.  The  MgH
photodissociation data turned out to be too weak to influence the temperature
structure of the atmosphere.

\begin{figure}
\plotone{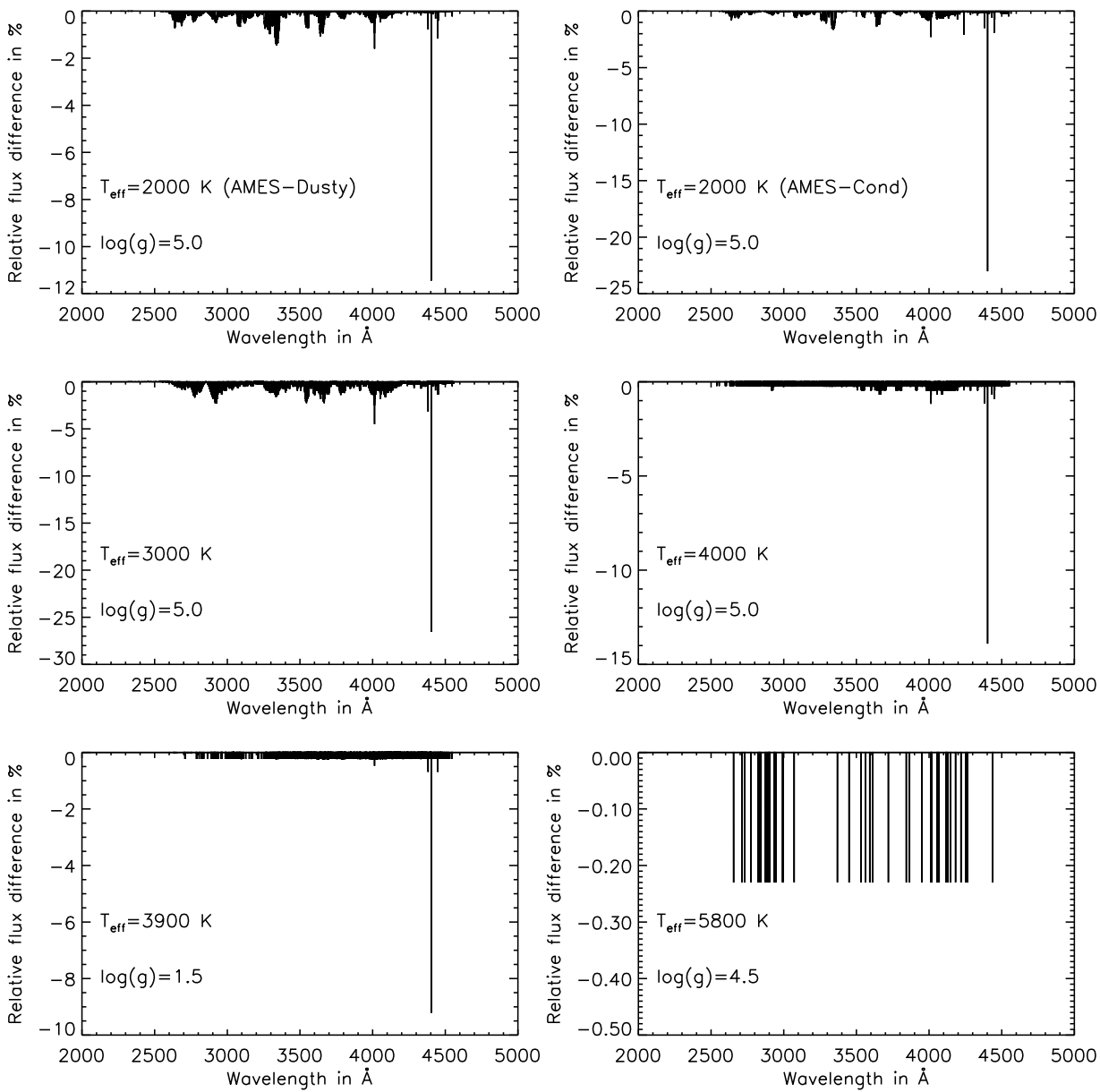}
\caption{Relative flux differences between model spectra after including
MgH photodissociation data. The parameters are indicated.
All models are AMES-Cond, except the top left one.
\label{difffig}}
\end{figure}

We calculated models with log($g$)=5.0 and effective temperatures of 2000~K,
3000~K, and 4000~K as typical M and L dwarf parameters.
To test if MgH is the missing opacity found by Short \& Lester (1994, 1996),
we calculated spectra  with log($g$)=4.5 and $T_{\rm eff}$=5800~K (as a
model to be close to a solar model) and with log($g$)=1.5 and $T_{\rm eff}$=3900
(as a typical K giant). We found changes in all spectra of typically
less then 2\% when including the MgH photodissociation 
data. This is demonstrated
in Figure \ref{difffig} where the relative flux differences are shown.
As can be seen, only the resonant dissociation transitions are strong
enough to be visible.
This is true for both the AMES-Dusty and AMES-Cond models although the
absolute flux changes are smaller in the AMES-Dusty model since there
is less flux in the optical due to strong dust opacity.
The differences in the solar model
are of the order of the numerical accuracy.
We also tested a model with metallicity $z=-1.0$ ($T_{\rm eff}$=3000~K
and log($g$)=5.0) and found the effect due to the bound-free opacities
to be practically unchanged compared to the models with
solar abundances.

\begin{figure}
\plotone{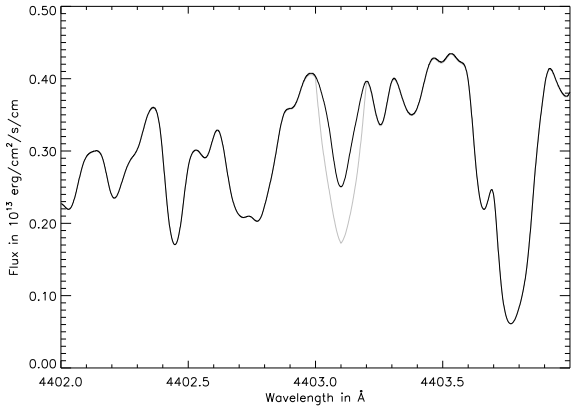}
\caption{Synthetic spectra at 0.001~\AA\ resolution in the
region of the bound-free resonance. The models have $T_{\rm eff}$=3000~K
and log(g)=5.0. The black line is without
bound-free absorption, the grey line is with bound-free absorption.
\label{specfig}}
\end{figure}

In Figure \ref{specfig} we show very high
resolution synthetic spectra in the region of the strongest resonance with
and without MgH bound-free data for the model with $T_{\rm eff}$=3000~K.
The feature is due to $B'\leftarrow X$ photodissociation from the $v''=0,
J''=43$ and $v''=1,J''=40$ levels of the $X$ state to quasi-bound levels of the
$B'$ state. However, the calculated wavelength of 4403~\AA~ is uncertain
due to uncertainties in the long-range behavior of both the $X$ and $B'$
potential energy curves. 
 
\section{Conclusion}
          
Photodissociation cross sections have been calculated for the
$A~^2\Pi\leftarrow X~^2\Sigma^+$ transition of $^{24}$MgH for 
all rotational transitions from the vibrational levels $v''=0-11$ and 
over the accessible wavelength range $\lambda\lambda~1770-4560$~\AA.
As predicted earlier by \cite{kir79}, using the Franck-Condon picture, 
photodissociation cross sections through the $A~^2\Pi$ are several orders 
of magnitude smaller than through the $B'~^2\Sigma^+\leftarrow X~^2\Sigma^+$ 
pathway. However, interesting features such as a large number of shape 
resonances, arising from rotational 
predissociation near thresholds, appear in the LTE photodissociation cross 
sections calculated for temperatures between 1000 and 5000~K using a 
Boltzmann distribution of the rovibrational levels of the $X~^2\Sigma^+$ 
state. Inclusion of the $^{24}$MgH $A\leftarrow X$ and $B'\leftarrow X$
photodissociation cross sections in atmosphere models of cool stars 
results in only minor changes of the computed spectra. 
Though it does not explain the missing opacity found by
Short \& Lester (1994, 1996),
the MgH bound-free opacities will have to be considered
for the analysis of high resolution spectroscopy at high S/N.
Further, as MgH is a trace molecule, the opacity due to photodissociation
may be more significant for molecules with larger abundances such as
H$_2$O.

\acknowledgments

This work was supported in part by NSF grants AST-9720704 and AST-0086246,
  NASA
grants NAG5-8425, NAG5-9222, and NAG5-10551 as well as NASA/JPL 
grant 961582 to the University
of Georgia. This work also was supported in
part by the P\^ole Scientifique de Mod\'elisation Num\'erique at ENS-Lyon.
Some of the calculations presented in this paper were performed on the IBM 
SP2 and the SGI Origin
of the UGA EITS, on the IBM SP ``Blue Horizon'' of the San Diego 
Supercomputer
Center (SDSC), with support from the National Science Foundation, and on 
the
IBM SP of the NERSC with support from the DoE.  We thank all these 
institutions
for a generous allocation of computer time.
This work was partially supported by the National Science Foundation 
through a grant for the Institute for Theoretical
Atomic \& Molecular Physics at Harvard University and Smithsonian 
Astrophysical Observatory.
P.F.W. and P.C.S. are grateful to Professor A. Dalgarno for useful 
discussions about this work. 

\appendix

\section{Appendix}

As a check on the accuracy of our numerical calculations, the sum 
rule for vibrational matrix elements proposed by \cite{ste72} is 
extended to rotational resolution. 
If $\{\chi_{v'J'},\chi_{k'J'}\} $ is 
a complete set of discrete, bound rovibrational 
wave functions $\chi_{v'J'}$ and 
energy normalized continuum wave functions $\chi_{k'J'}$, and $D$ an operator, then, from 
the equality case of the Schwarz inequality, we obtain
\begin{equation}
<\chi_{v''J''}\vert D^{2}\vert\chi_{v''J''}>=\frac{1}{2J''+1}
[{\cal D}^{bound}_{v''J''}
+{\cal D}^{free}_{v''J''}],
\end{equation}
with, for transitions between bound states of the $X~^2\Sigma^+$ and 
$A~^2\Pi$ electronic states,
\begin{eqnarray}
{\cal D}^{bound}_{v''J''}&=
&\sum_{v',J'}S_{J'}\vert<\chi_{v'J'}\vert D\vert\chi_{v''J''}>\vert^{2}
\nonumber \\
&=&\sum_{v'}\{S_{J'}^{R}\vert<\chi_{v'J''+1}\vert D\vert\chi_{v''J''}>\vert^{2}
\nonumber \\
& &~~~~~ +S_{J'}^{Q}\vert<\chi_{v'J''}\vert D\vert\chi_{v''J''}>\vert^{2}
+S_{J'}^{P}\vert<\chi_{v'J''-1}\vert D\vert\chi_{v''J''}>\vert^{2}\}
\end{eqnarray}    
and, for transitions from bound states of the $X~^2\Sigma^+$ electronic state to 
continuum states of the $A~^2\Pi$ electronic state,
\begin{eqnarray}
{\cal D}^{free}_{v''J''}&=&\int_{E_{k'J'}}S_{J'}
\vert<\chi_{k'J'}\vert D\vert\chi_{v''J''}>\vert^{2}
dE_{k'J'} 
\nonumber \\
&=&\int_{E_{k'}}\{S_{J'}^{R}\vert<\chi_{k'J''+1}\vert D\vert\chi_{v''J''}>\vert^{2}
\nonumber \\
& &~~~~~ +S_{J'}^{Q}\vert<\chi_{k'J''}\vert D\vert\chi_{v''J''}>\vert^{2}
+S_{J'}^{P}\vert<\chi_{k'J''-1}\vert D\vert\chi_{v''J''}>\vert^{2}\} dE_{k'},
\end{eqnarray}
where we have introduced the H\"{o}nl-London factors $S_{J'}^{P}$, $S_{J'}^{Q}$ and 
$S_{J'}^{R}$ defined in equation (\ref{hlfac}) for the $P$-, $Q$- and 
$R$-branches, respectively.

For absorption from the rotational levels of the $X~^2\Sigma^+~(v''=0)$ vibrational 
state, the sum rule,
using the photodissociation cross section presented here and
the bound-bound oscillator strengths of \cite{wec02b},
 was satisfied to within an error of $0.0008\%$ for $J''=0$ 
(${\cal D}^{f/b}_{0~0}={\cal D}^{free}_{0~0}/ 
{\cal D}^{bound}_{0~0}=8\times 10^{-6}$)
up to a maximum error of $0.032\%$ for the highest-lying 
rotational level, $J''=44$ 
(${\cal D}^{f/b}_{0~44}=6\times 10^{-4}$).
As the vibrational and rotational quantum numbers 
of the ground electronic state increase, 
the error increases, for example, with errors of $0.0078\%$ 
for $v''=2$ and $J''=0$ 
(${\cal D}^{f/b}_{2~0}=8\times 10^{-5}$)
and $7.08\%$ for the topmost rotational level 
of $v''=2$, $J''=39$
(${\cal D}^{f/b}_{2~39}=7\times 10^{-2}$).
           
As a remark, let us note that the present sum rule is aimed at checking the 
accuracy of the numerical calculations and in no way reflects the validity of the 
potential surfaces and the dipole transition moment. 



\clearpage

\onecolumn

\clearpage
\twocolumn

\clearpage

\clearpage
\onecolumn

\clearpage

\clearpage

\twocolumn

\end{document}